\documentclass[12pt]{article}

\usepackage{indentfirst}
\usepackage{amsmath}
\usepackage{hyperref}
\usepackage{amssymb}
\usepackage{amsfonts}
\usepackage{amscd}
\usepackage{amsbsy}
\usepackage{amsthm}
\usepackage{latexsym}
\usepackage{graphicx,color} 
\usepackage{booktabs}
\usepackage{verbatim}

\def\nn{\nonumber}       

\renewcommand{\vec}[1]{{\bf #1}}

\def\beq{\begin{eqnarray}}
\def\eeq{\end{eqnarray}}
\def\ln{\,\mbox{ln}\,}

\def\diag{\,\mbox{diag}\,}

\def\al{\alpha}
\def\be{\beta}

\def\ga{\gamma}
\def\de{\delta}
\def\vp{\varepsilon}

\def\ze{\zeta}

\def\la{\lambda}
\def\na{\nabla}
\def\pa{\partial}

\def\si{\sigma}
\def\om{\omega}

\def\La{\Lambda}

\def\Om{\Omega}

\renewcommand{\vec}[1]{{\bf #1}}

\def\CC{cosmological constant}

\def\EA{effective action}

\begin{document}

\begin{center}

{\large\bf
Constraints from observational data for a running cosmological
constant and warm dark matter with curvature}
\vskip 6mm

\textbf{Jhonny A. Agudelo Ruiz}$^{a,c}$
\footnote{E-mail address: \ jaar@cosmo-ufes.org},
\ \
\textbf{J\'ulio C. Fabris}$^{a,b}$
\footnote{E-mail address: \ julio.fabris@cosmo-ufes.org},
\\
\textbf{Alan M. Velasquez-Toribio}$^{a}$
\footnote{E-mail address: \  alan.toribio@ufes.br},
\ \
\textbf{Ilya L. Shapiro}$^{c}$ 
\footnote{E-mail address: \ shapiro@fisica.ufjf.br}
\vskip 3mm

$^{a}$
\textsl{N\'ucleo Cosmo-UFES \& PPGCosmo, Departamento de
F\'isica,
\\
Universidade Federal do Esp\'irito Santo, Vit\'oria,
29075-910, ES, Brazil}
\vskip 2mm

$^{b}$
\textsl{National Research Nuclear University MEPhI,
Kashirskoe sh. 31,
\\
Moscow 115409, Russia}
\vskip 2mm

$^{c}$
\textsl{Departamento de Física, ICE,
Universidade Federal  de Juiz  de  Fora,
\\
Juiz de Fora, 36036-100, MG, Brazil}
\vskip 2mm



\end{center}
\vskip 6mm

\centerline{{\bf Abstract}}

\begin{quotation}
\noindent
It is known than the inclusion of spatial curvature can modify the
evolution of matter perturbations, and affect the Large Scale Structure
(LSS) formation. We quantify the effects of the non-zero space
curvature in terms of LSS formation for a cosmological model
with a running vacuum energy density and a warm dark matter
component. The evolution of density perturbations and the modified
shape of its power spectrum are constructed and analyzed.
\vskip 3mm

\noindent
{\sl Keywords:}
\ Large scale structure, running cosmological constant,
warm dark matter, matter power spectrum
\vskip 2mm

\end{quotation}

\newpage


\section{Introduction}

The fiducial theory in the modern cosmology is based on general
relativity (GR) and the concordance model ($\Lambda$CDM),
where the main observables, including the large scale properties of
the universe, are explained in terms of the dark energy (DE) and
dark matter (DM) components
\cite{Planck2018Parameters,Boss2014}.

In $\Lambda$CDM, the global expansion of the Universe caused
by this DE is provided by the inclusion of the most natural and simple
alternative, which is the cosmological
constant (CC), characterized by the equation of state $\om_\La=-1$,
or $P_\La=-\rho_\La$. Independent of the difficult CC problem
\cite{Weinberg89}, the presence of this constant is also
a consistency requirement from the anthropic \cite{Weinberg-87antrop}
and quantum field theory viewpoints \cite{CC-nova,PoImpo}. In the
phenomenological framework the CC is a very successful concept,
regardless of some deviations in the equation of state of the DE from
$\om_{DE}=-1$ can not be ruled out (see e.g.
\cite{Sahni2000,Peebles2003,Sahni-Star}). On the other hand, DM
also emerges as a necessary element
for explaining the observational evidence about the large scale
structure (LSS) formation, baryon acoustic oscillations (BAO) and
cosmic microwave background (CMB) anisotropies
\cite{Bertone2005,Tegmark2004,Planck2018CMB}.

However, despite of the mentioned success of $\Lambda$CDM,
there are still some discrepancies between different cosmic data sets
and measurements, which opens up the door to alternatives theories of
gravity or new generalizations and extensions of this standard model
(see e.g. \cite{Capozziello2011} for the review). Thus, in order to
alleviate these tensions is necessary to look for alternatives
descriptions for the gravitational sector or for the cosmic
components.

Due to the mentioned theoretical importance and the experimental
necessity of the CC, it would makes sense then to explore the
possibility that it may be slowly varying, in particular, as
consequence of the low-energy quantum corrections. In general,
such quantum corrections can be consistently described by the
renormalization group running within the semiclassical theory
(see e.g. \cite{book}) or in quantum gravity \cite{TV90,BrTibUAE}.
One can note that the quantum gravitational running at low energies
(in the IR) is well-defined but phenomenologically almost irrelevant
\cite{BrTibUAE}, while the semiclassical running of the density of
cosmological constant $\rho_\La$ in the IR can be formulated only
phenomenologically and characterized by a single free parameter
$\nu$ \cite{CC-nova,DCCrun}. The models based on such a running
were originally developed in \cite{CC-fit,CCwave} (background
and cosmic perturbations)
and \cite{CC-Gruni,CCG}, producing many interesting developments
(see e.g. \cite{Sola2013-1,PerTam,BMS2} and references therein).

On the other hand, describing DM component in terms of particles or
fields also represents an open question. Besides the theoretical and
experimental difficulties in detecting DM, there is a possibility to
assume that DM is warm (WDM) instead of cold (CDM). It is
known that the relativistic warmness of DM can change the global
dynamic of the universe and provide certain phenomenological
advantages \cite{Hannestad2000,Bode2001,Viel2005}. The full standard
description of WDM implies the use of the Boltzmann equation.
However,  the problem can be greatly simplified using the reduced
relativistic gas model (RRG) \cite{Sakharov1966, FlaFlu}. 
The main point of the RRG is that it assumes the same kinetic energy
for all particles of relativistic gas. Such an artificial ideal relativistic
gas model has a very simple equation of state with the unique free
parameter $b$, characterizing its warmness. On the other hand, this
equation of state closely reproduce similar equation in the J\"uttner
model \cite{Juttner1911}, based on the relativistic Maxwell
distribution. These two features enable one to use RRG, e.g., for the
simplified phenomenological description of WDM
\cite{Sobrera2009,Fabris2012testing,Hipolito2017}. The simplicity
of RRG is especially welcome in the theories with
natural complications, such as the cosmological models with running
parameters. As far as most interesting models of running $\rho_\La$
(see e.g. \cite{CC-fit} and \cite{CCwave}) are consistently applicable only
at high energy scale, we need to formulate them in the framework
of early universe, when the DM is supposed to have more warmness
than today. Then the RRG becomes a useful tool that enables to get
the main features of the model with the reduced amount of numerical
calculations and more clear physical understanding of the results.
For this reason, in the recent work \cite{RGGRcosm} we have
started the exploration of the running $\rho_\La$ and RRG model
for the early Universe.

In the present work, we continue the previous development of
\cite{RGGRcosm} and consider the consequences of running
cosmological constant using the approach of \cite{CC-fit} and
\cite{CCwave}, which is consistent only in the early Universe.
Different from the later epochs, in the epoch soon after inflation,
the creation of at least the Standard Model particles from the
vacuum (see e.g. \cite{ZeldStar71}, \cite{DobMar} and references
therein), is not suppressed by the low energy density of the
gravitational field of the cosmological background. At the same
time, the running of the cosmological constant density in the
high-energy regime is the phenomena which may leave
observational traces in the late universe. This is the subject of
the study in Ref.~\cite{RGGRcosm}. In this paper we make the next step and quantify the
effects of the non-zero spatial curvature in the model with running
$\rho_\La$ and the WDM contents described by RRG model in the
early universe. Our purpose is to evaluate the effect of spatial
curvature on some observables in the context of LSS formation as the matter power spectrum. Technically, our
purpose is to evaluate the constraints on the free parameters $\nu$
and $b$ in the presence of $\Omega^{0}_{k}$, using SNIa and
DR11 cosmic data-sets \cite{scolnic,anderson}.
Indeed, it is interesting to include curvature in the model of
Ref.~\cite{RGGRcosm}, and not only for the sake of generality.
In the last years, there was an intensive discussion of the
observational constraints on the space geometry, including the
curvature of the universe. For example, some observational results
including SNIa, H, BAO, QSO, etc have shown statistical consistency
with a closed curvature universe (see e.g. the references
\cite{ooba}-\cite{handleyy}).
Thus, it looks natural to include consideration of space curvature
in the model with the running cosmological constant.

The paper is organized as follows. In the next Sec.~\ref{s2}, the
system of equations for the background cosmological model are
derived, where it is found a closed expression for the expansion
rate. In Sec.~\ref{s3}, the density
perturbations are obtained and the system of equations for the
density contrasts solved numerically in order to reconstruct the
matter power spectrum for our model. In Sec.~\ref{s4}, the
constraints for the free parameters of our model in the presence
of spatial curvature are found and discussed. Finally, in
Sec.~\ref{s5}, we draw our conclusion and discuss some perspectives.

\section{Background solution}
\label{s2}

As it was mentioned already,
we want to consider the presence of spatial curvature and to analyze
the consequences for a cosmic epoch after recombination, this is, in
a matter-dominated (MD) universe with the aim of finding some
changes in the matter power spectrum of matter and some new
constraints for the free parameters $\nu$ and $b$ with respect to
$\Omega^{0}_{k}$. For the sake of generality,
we hold $\omega$ in all the expressions, but when starting the
numerical estimates in Sec. \ref{s4}, we shall set $\omega=0$, as
expected for a usual matter (in what follows we call it baryonic)
component in a MD universe after recombination. Thus, the
equations of state for the baryonic matter, running CC density
and WDM are defined by the relations \cite{FlaFlu} (see
also \cite{RRG-anisotrop} and \cite{RRG-Leo} for alternative
derivations)
\beq
\label{6}
 &&
 p_b=\omega\rho_b,
 \qquad
 \mbox{and}
 \qquad
 p_{\Lambda}=-\rho_{\Lambda},
\\
\label{7}
&&
p_{dm}\,=\, \frac{\rho_{dm}}{3}
\Big[1 -  \Big(\frac{m c^2}{\vp} \Big)^2 \Big]
\,=\, \frac{\rho_{dm}}{3}\,(1 - s),
\eeq
where
\beq
\label{8}
s= \frac{\rho_d^2}{\rho_{dm}^2},
\qquad\qquad
\vp = \frac{mc^2}{\sqrt{1-\be^2}},
\qquad\qquad
\rho_{dm}=n\vp,
\eeq
$\vp$ and $n$ are the kinetic energy of the individual particle and
the concentration of these particles, and $\rho_d=nmc^2$ is the density
of the rest energy, with the scaling rule
\beq
\label{9}
\rho_d(z) = \rho_{d}^0 (1+z)^3.
\eeq

The mathematical description of the model is based on GR, with a
non-zero spatial curvature, energy exchange between cosmological
constant density and baryonic matter, and adiabatically expanding
ideal gas of WDM, described by RRG. In this way, we arrive at
the following system of equations:
\beq
\label{1}
&&
H^2 (z) \,=\, \frac{\kappa^2}{3} \big[
\rho_\La (z) + \rho_b (z) + \rho_{dm} (z)\big]
+ H_{0}^{2}\Omega_{k}^{0}(1+z)^{2}
\\
\label{2}
&&
\rho_b' - \frac{3(1+w)}{1+z} \, \rho_b = - \rho_\La',
\\
\label{3}
&&
\rho_{dm}' = \frac{(4-s)}{1+z} \, \rho_{dm}.
\eeq
Eq.~(\ref{1}) is the Friedmann equation with the space curvature term,
compared to the similar equation in \cite{RGGRcosm}, where one can
find more details. Eqs.~(\ref{2}) and ~(\ref{3}) describe the
conservation of the energy-momentum tensor for the baryonic matter
$\rho_b$ and the running vacuum $\rho_{\Lambda}$, and the
conservation law for the WDM component modelled as a RRG, as
explained above.

In what follows we will need the total energy-momentum tensor,
that is given by the sum of the baryonic, vacuum and WDM parts,
\beq
\label{4}
T^{\mu}_{\nu}\,=\,{L}^{\mu}_{\nu}+{M}^{\mu}_{\nu},
\eeq
where
\beq
&&
{L}^{\mu}_{\nu}
\,=\,
(1+\omega)\rho_{b}U^{\mu}U_{\nu}
- (\omega\rho_{b}-\rho_\Lambda)\delta^{\mu}_{\nu},
\nn
\\
&&
{M}^{\mu}_{\nu}
\,=\, \frac{4-s}{3}\rho_{dm}V^{\mu}V_{\nu}
- \frac{1-s}{3}\rho_{dm}\delta^{\mu}_{\nu},
\label{5}
\eeq
with the associated 4-velocities $U^\mu$ and $V^\mu$.

On the other hand, from the quantum field theory (QFT) perspective
we know that the running of vacuum energy density is determined by
the possible quantum contributions of massive field in the vacuum
effective action. One can parameterize the running of $\rho_\La$ in
terms of a free parameter $\nu$ \cite{CC-nova,PoImpo,CC-fit}, as
\beq
\label{10}
\frac{d\rho_\La}{dz} \,=\,\frac{3\nu}{8\pi G}\frac{dH^{2}}{dz}
\eeq
or, equivalently, as
\beq
\label{11}
\rho_\La \,=\, \rho^0_\La \,+\,
\frac{3\nu}{8\pi G}\, \big( H^2 - H_0^2),
\eeq
where the sign of $\nu$ indicates whether bosons or
fermions dominate in the running \cite{CC-fit}. Let us note that
there is nowadays an extensive literature on the covariant
realization of this and similar forms of running (see e.g.
\cite{BWD} and refences therein).

As it was recently discussed in \cite{RGGRcosm}, the running of the
cosmological constant can be compatible with the energy transfer
from vacuum to matter only in the early Universe, there the intensity
of the gravitational background metric (characterized by the Hubble
parameter $H$) is sufficient for producing the normal particles,
(let us remember that we call it baryonic matter). For instance, this
is possible in the reheating period after inflation. e.g. for the
phenomenologically successful Starobinsky (or $R^2$) inflation
\cite{star,star83}, the typical energy at the end of inflationary period
is about $10^{13}\,GeV$. This and even much lower energy scales
is certainly sufficient to produce normal particles with the masses
below $10^{3}\,GeV$. Moreover, these particles would have
kinetic energies many orders above their masses and, therefore,
have the equation of state very close to the one of radiation.
However, for some models of dark matter, e.g. the ones based on
the Grand Unification remnants, at this energy scale the production
of the corresponding particles is impossible. In this physical
situation dark matter is a warm but ideal gas of particles which do
not interact with the rest of the world, except gravitationally.

The last observation concerns the equation of state of the
cosmological constant term which is actually non-constant,
according to Eq.~(\ref{11}). In order to address this issue,
let us remember that the quantum or semiclassical corrections
which are behind the running of any quantum field theory
parameter, are typically non-local and rather complicated. How
can we separate those terms in the vacuum effective action, that can
be attributed to the \CC \ with quantum contributions? The solution
that looks reasonable at least for the cosmological applications is
that \CC \ terms in the \EA \  should scale like the classical \CC \
under the global transformation $g_{\mu\nu}\to g_{\mu\nu}e^{2\la}$,
where $\,\la=const$. For instance, such terms as \cite{apco}
\beq
&&
\int d^4\sqrt{-g}\, R \,\frac{1}{\Box^2}\, R,
\qquad
\int d^4\sqrt{-g}\, R_{\mu\nu}
\,\frac{1}{\Box^2}\, R^{\mu\nu}\,,
\quad
\mbox{or}
\quad
\int d^4\sqrt{-g} \,R_{\mu\nu\al\be}
\,\frac{1}{\Box^2 }\, R^{\mu\nu\al\be},
\mbox{\qquad}
\label{nonlocCC}
\eeq
and many other similar structures, belong to this group and could be
used as toy models for quantum corrections to the \CC. For a
constant scaling these terms transform exactly as the \CC \ term.
For the FRW (homogeneous and isotropic) metric, the difference
with the \CC \ is proportional to the derivatives of the non-constant
scaling parameter $\si(t)$, after we replace $\,\la \to \si(t)$. That is
why these terms provide a cosmological models with the equivalent
of a slowly varying \CC \ \cite{MM} (see also further references
therein). Thus, as far as the global scaling of the hypothetical
terms responsible for the running (\ref{11}) is the same as for the
cosmological constant, the equation of state for the corresponding
term is assumed to be the same as in the constant $\rho_\La$ case
\cite{CC-fit}.

The set of equations formulated above, is appropriate for a
simple albeit reliable description of the phase of the Universe
with the running cosmological constant and the energy exchange
with the matter sector.
To solve the system of four equations \eqref{1}-\eqref{3} and
\eqref{10}, note first that in our model the WDM component is
decoupled, so we can solve its conservation law directly to get
\beq
\label{12}
\Om_{dm}(z)
\, = \,
\frac{\Om_{dm}^0(1+z)^3}{\sqrt{1+b^2}} \, \sqrt{1+b^2(1+z)^2},
\eeq
where $\Om^0_{dm}$ is the DM density in the present-day Universe
\footnote{Here and from now on we use the notations
$\Omega_i (z)=\rho_i (z)/\rho^{0}_c\,$, where
$\rho^0_c \,=\, 3 H_0^2/8 \pi G$. It is easy to see that this is
the density relative to the critical density at $a_0$, that means
nowadays, and not to the time-dependent density.} and
\beq
\label{13}
b = \frac{\be}{\sqrt{1-\be^2}}
\eeq
measures the warmness. E.g. in the nonrelativistic limit $\,\be \ll 1$
we have $\,b \sim \be$, hence $b \approx  0$ means that the DM
contents is ``cold''. Eq.~(\ref{12}) shows the scaling rule for the
model. Being taken alone, the RRG provides a natural and simple
interpolation between the radiation- and matter-dominated
cosmological solutions \cite{Sakharov1966,FlaFlu}.

Using \eqref{10} in \eqref{1} we get
\beq
\label{14}
\frac{d\rho_\Lambda}{dz}=\frac{\nu}{1-\nu}
\left[\frac{d\rho_b}{dz}+\frac{d\rho_{dm}}{dz}
+ \frac{6H_{0}^{2}}{\kappa^2}\,\Omega_{k}^{0}(1+z)\right]
\eeq
and, substituting this result in \eqref{2}, we at the simple differential
equation for $\rho_b$,
\beq
\label{15}
\frac{d\rho_b}{dz} - \frac{\ze}{1+z} \rho_b  \,=\,  - \nu\, \frac{d\rho_{dm}}{dz}-\frac{6\nu}{\kappa^2}
\, H_{0}^{2}\Omega_{k}^{0}(1+z),
\eeq
where $\rho_{dm}$ is a solution to (\ref{3}) and the useful new
parameter is
\beq
\label{18}
\ze  \,=\,  3(1+w)(1-\nu).
\eeq
The solution to Eq.~(\ref{15}) is given by
\beq
\Om_b (z)
&=&
C_0 (1+z)^\ze-\frac{2\nu}{2-\zeta}\Omega_{k}^{0}(1+z)^{2}
\nn
\\
&-&
\frac{\nu \Om_{dm}^0(1+z)^3}{\sqrt{1+b^2}}
\Big[ \sqrt{1+b^2 (1+z)^2}
+ \frac{\zeta}{3-\zeta}\,\, {}_{2} F_1 (\al, \be; \ga; Z) \Big],
\mbox{\qquad}
\label{16}
\eeq
where
\beq
\label{17}
C_0 = \Om_{b}^0+\frac{2\nu}{2-\zeta}\Omega_{k}^{0}
+ \frac{\nu  \Om_{dm}^0}{\sqrt{1+b^2}}
 \Big[ \sqrt{1+b^2} +  \frac{\zeta}{3-\zeta } \,\,
{}_ 2F_1 (\al, \be; \ga;-b^2) \Big],
\eeq
and \ ${}_2 F_1(\al, \be; \ga; Z)$ is the hypergeometric function
defined as
\beq\label{19}
{}_2 F_1(\al, \be; \ga; Z) = \sum_{k=0}^\infty \frac{(\al)_k (\be)_k}{(\ga)_k}
\,\frac{Z^k}{k!},
\eeq
and $(\al)_k\,$ is the Pochhammer symbol. In our case
\beq
\label{20}
\al = -\frac12,
\qquad
\be = \frac{3-\ze}{2},
\qquad
\ga = \frac{5-\ze}{2}
\qquad
\mbox{and}
\qquad
Z = -b^2(1+z)^2.
\eeq
\begin{figure}[t!]
\centering
\includegraphics[scale=0.6]{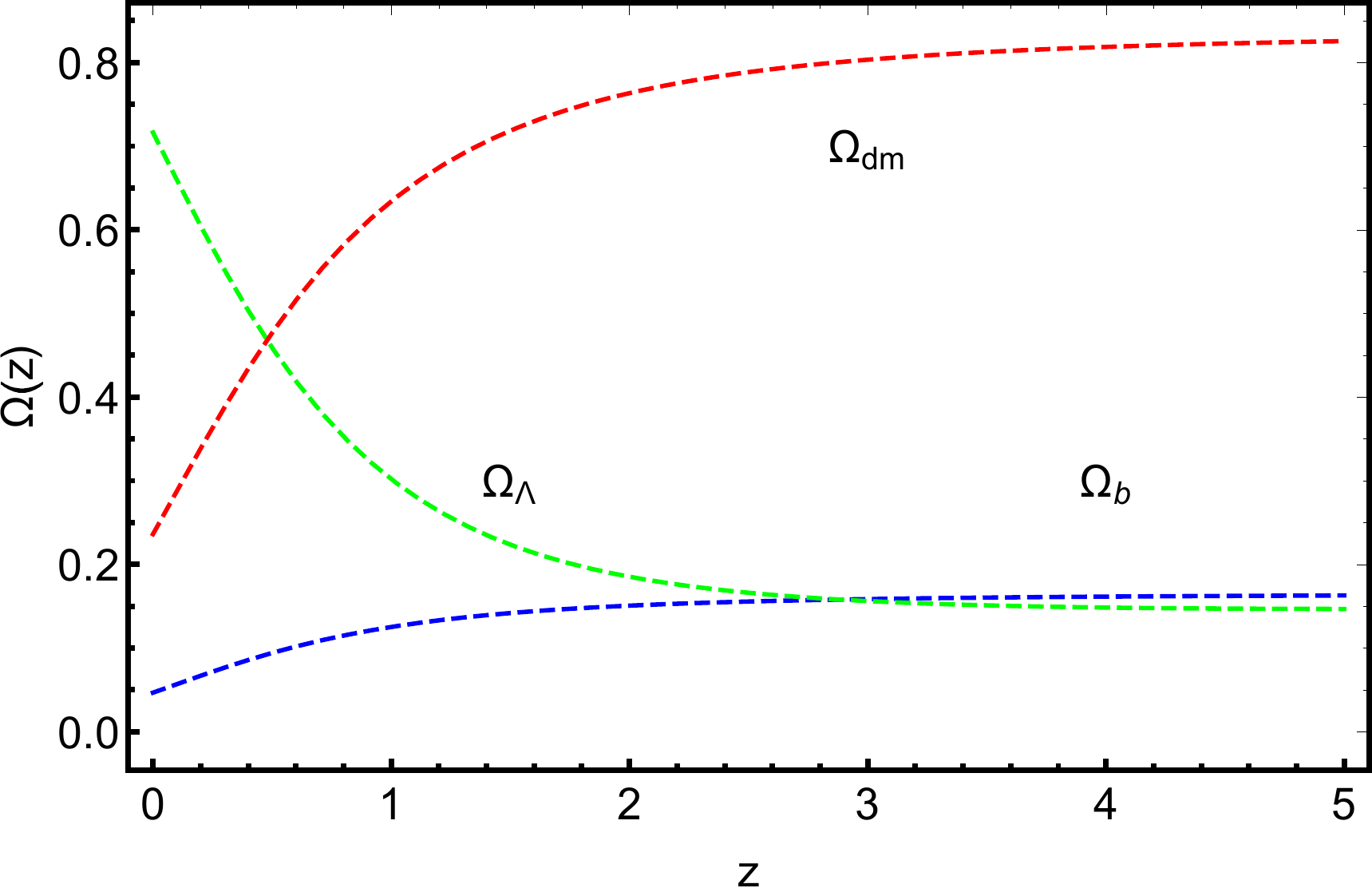}
\caption{\footnotesize
Dynamic evolution of the three
energy-matter components of the Universe. One can observe
the WDM domination and the domination of the running
cosmological constant part for big and small
redshifts $z$, respectively. The growing of DE as a consequence
of baryons decay, it is also observed.}
\label{fig1}
\end{figure}

The solution for $\rho_\Lambda$ can be found by integrating \eqref{12},
\beq
\label{21}
\Om_\La(z) =  B_0 + \frac{\nu}{1-\nu} \, \left[\Om_r(z)
+ \Om_{dm}(z)+\Omega_{k}^{0}\,z(z+2)\right],
\eeq
where
\beq
\label{22}
B_0 \,=\,
\Om_{\La}^0 - \frac{\nu}{1-\nu} \, \big(\Om_{r}^0 + \Om_{dm}^0\big).
\eeq
Finally, for the square of the Hubble parameter we find
\beq
\left(\frac{H(z)}{H_{0}}\right)^{2}
&=&
1+\left(\Omega_{b}^{0}+\frac{2\nu\Omega_{k}^{0}}{2-\zeta}\right)
\left[\frac{(1+z)^{\zeta}-1}{1-\nu}\right]
+ \Omega_{k}^{0}(z^{2}+2z)
\left[1-\frac{\nu\zeta}{(1-\nu)(2-\zeta)} \right]
\nn
\\
&+&
\frac{\Omega_{dm}^{0}}{1-\nu}\left\{\left[\nu+\frac{\nu\zeta}{3-\zeta }
\,\,
\frac{{}_ 2F_1 (\al, \be; \ga;-b^2)}{\sqrt{1+b^2}}\right]
(1+z)^{\zeta}-1\right\}
\nn
\\
&+&
\frac{\Om_{dm}^0(1+z)^3}{\sqrt{1+b^2}}
\Big[ \sqrt{1+b^2 (1+z)^2}
- \frac{\nu\zeta}{(1-\nu)(3-\zeta)}\,\, {}_{2} F_1 (\al, \be; \ga; Z) \Big].
\label{23}
\eeq
In the limits of $\nu\to 0$ and $b\to 0$ we recover the standard
$\Lambda$CDM model and in the case of $\omega=0$ and
$\Omega_{dm}^{0}=0$ (that is, without WDM), we recover exactly
the result presented in \cite{CCwave}.  In Figure \ref{fig1} is
presented the cosmic evolution of the relative energy densities for
normal matter, running vacuum and warm dark matter with respect
to redshift $z$.

\section{Cosmic perturbations}
\label{s3}

Let us consider the cosmological perturbations in the RRG with
running $\rho_\La$, following the approach developed in
Refs.~\cite{CCwave} and \cite{Sobrera2009}. The first observation
is that the perturbation of the WDM pressure should be derived
from the equation of state \eqref{7},
\beq
\label{25}
\de p_{dm}
\,=\, \frac{\de \rho_{dm}}{3}
\Big[ 1 - \Big(\frac{mc^2}{\vp} \Big)^2 \Big]
\,=\, \frac{\de \rho_{dm} (1-s)}{3}.
\eeq
Taking the main feature of the RRG, this means that perturbations
satisfy the same relation between pressure and energy density that
the background quantities. The metric perturbations are defined by
\beq
\label{26}
g_{\mu\nu}=\bar{g}_{\mu\nu}+h_{\mu\nu},
\eeq
with the associated Christoffel symbol
\beq
\label{27}
\Gamma^{\rho}_{\mu\nu}
= \bar{\Gamma}^{\rho}_{\mu\nu}
+ \frac{1}{2}g^{\rho\lambda}
\left(\partial_{\mu}h_{\lambda\nu}
+\partial_{\nu}h_{\lambda\mu}
-\partial_{\lambda}h_{\mu\nu}
-2h_{\lambda\kappa}\bar{\Gamma}^{\kappa}_{\mu\nu}\right),
\eeq
where the background quantities are marked by bars. E.g., the
background metric has the form
\beq
\label{28}
\bar{g}_{\mu\nu}\,=\,\diag\big\{1,-a^{2}(t)\delta_{ij}\big\}.
\eeq
In the synchronous gauge $h_{0\mu}=0$, the $(00)$ component
of the Ricci tensor is
\beq
\label{29}
R_{00}=\bar{R}_{00}+\delta R_{00},
\eeq
where
\beq
\label{30}
\delta R_{00}
\,= \,\frac{1}{2}\,\dot{h}+Hh
\qquad
\mbox{and}
\qquad
h\,=\,\frac{\pa}{\pa t} \Big( \frac{h_{ii}}{a^2}\Big).
\eeq
Perturbing the Einstein equations, we obtain
\beq
\label{31}
\delta R^{\mu}_{\nu}
\,=\,8\pi G
\left(\delta T_{\nu}^{\mu}
- \frac{1}{2}T g^{\rho\mu}\delta g_{\rho\nu}
- \frac{1}{2}\delta T \delta^{\mu}_{\nu}\right),
\eeq
where the total energy-momentum tensor
$T_{\mu\nu}$ and its trace $T$ are given by Eq.~\eqref{4}.

It proves useful introducing the quantities
\beq
\label{32}
f_1 (z) = \frac{\rho_b (z)}{\rho_t (z)},
\qquad
f_2 (z) = \frac{\rho_\La (z)}{\rho_t (z)},
\qquad
f_3 (z) = \frac{\rho_{dm} (z)}{\rho_t (z)},
\eeq
and
\beq
\label{33}
g(z) = \frac{2\nu H(z)}{3H^{2}(z)
- 3H_{0}^{2} \Omega_{k}^{0}(1+z)^{2}},
\eeq
where $\rho_t$ is the total energy density. Thus, we arrive at the
$00$-component of the linearized Einstein equations,
\beq
\label{34}
h' - \frac{2h}{1+z}
\,=\, - \frac{2\nu}{(1+z)g} \big[
(1+3w)f_1 \de_b - 2 f_2 \de_\La + (2-s)f_3 \de_{dm}
\big],
\eeq
where
\beq
\label{35}
\de_i = \frac{\de \rho_i}{\rho_i}\,.
\eeq
is the relative density variation.

It is useful to denote
\beq
v = f_1 \na_i (\de V^i)
\qquad
\mbox{and}
\qquad
u = f_3 \na_i (\de U^i)
\eeq
the divergences of the peculiar velocities. The time and spatial
components of  the perturbations satisfy the linear equations
\beq
&&
\de_b'
\,+\, \Big[
\frac{f_1'}{f_1} - \frac{3(1+w)f_2}{1+z}
+ \frac{(1-s-3w)f_3}{1+z}
\Big] \de_r
\,-\,
\frac{1+w}{(1+z)H}
\Big(\frac{v}{f_1} - \frac{h}{2} \Big)
\nn
\\
&&
\,\,\quad
=\, -\,\frac{1}{f_1}(\de_\La f_2)'
\,-\, \frac{3(1+w)f_2}{1+z}
\Big[ 1 + \frac{(4-s)f_3}{3(1+w)f_1} \Big]
\de_\La,
\label{36}
\\
&&
v' + \frac{[3(1+w)f_1+(4-s)f_3-5]}{1+z} \, v
\,=\,
\frac{k^2(1+z)}{(1+w)H} \left(f_2 \de_\La - w f_1 \de_b \right),\label{37}
\\
&&
\de'_{dm}
+ \Big\{
\,\frac{f_3'}{f_3}
+ \frac{3(1+w) f_1 + (s-4) (f_1 + f_2)}{1+z}
\Big\} \de_{dm}
\,+\, \frac{4-s}{3H(1+z)} \Big( \frac{h}{2} - \frac{u}{f_3} \Big)
\,=\, 0,
\mbox{\quad}
\mbox{\qquad}
\label{38}
\\
&&
u' + \Big[
\frac{3(1+w)f_1+ (4-s)f_3 -5}{1+z}
 - \frac{s'}{4-s}
\Big] u
+ \frac{k^2(1+z)f_3}{H} \Big( \frac{1-s}{4-s} \Big) \de_{dm}
\,=\, 0.
\mbox{\qquad\quad}\label{39}
\eeq
Perturbing Eq.~\eqref{11}, we get
\beq
\label{40}
\de_\La = \frac{g}{f_2} \Big(\frac{v}{f_1}-\frac{h}{2} \Big).
\eeq
It is easy to note that this equation is not dynamical, representing
a constraint that should be used in other equations. Using \eqref{40}
in Eqs.~\eqref{34}, \eqref{36} and \eqref{37}, and rewriting the
equations in the Fourier space, we arrive at the equations
\beq
\label{41}
&&
h' + \frac{2(\nu-1)}{1+z}\,h = \frac{2\nu}{1+z} \Big[
\frac{2v}{f_1}-(1+3w)\frac{f_1}{g} \de_b
- (2-s)\frac{f_3}{g} \de_{dm}\Big],
\\
\label{42}
&&
\de_b' + \Big[
\frac{f_1'}{f_1} - \frac{3(1+w)f_2}{1+z}
+ \frac{(1-s-3w)f_3}{1+z}
\Big] \de_b
= \frac{1}{f_1}\Big(\frac{gh}{2} -\frac{gv}{f_1}  \Big)'
\mbox{\qquad}
\nonumber
\\
&&
\qquad
+\, \frac{1+w}{1+z}
\Big[ 3g + \frac{(4-s)g f_3}{(1+w)f_1} - \frac{1}{H} \Big]
\Big(\frac{h}{2} -\frac{v}{f_1}  \Big),
\\
\label{43}
&&
v' + \Big\{\frac{[3(1+w)f_1+(4-s)f_3-5]}{1+z}
- \frac{k^2 g (1+z)}{(1+w)H f_1} \Big\} v
\nn
\\
&&
\qquad
=\,
-\frac{k^2 g(1+z)}{2(1+w)H} \Big( h + \frac{2w f_1}{g}\,\de_b \Big).
\eeq
Thus, the closed set of perturbation equations is given by
\eqref{38}, \eqref{39}, \eqref{41}, \eqref{42} and \eqref{43}.

\section{Some observational constraints}
\label{s4}

Let us use the equations calculated above and some of the
available observational
data to constrain the parameters of our model, including $\nu$, $b$
and $\Omega_{k}^{0}$.

\subsection{Supernovae Ia}

In this subsection we will use the data from Supernovas Ia called
``Pantheon'' sample \cite{scolnic}, which is the largest combined
sample of SNIa and consists of $1048$ data with the redshifts in
the range $0.01 < z < 2.3$. It is a collection of the SNe Ia,
discovered by the Pan-STARRS1 (PS1) Medium Deep Survey
and SNe Ia from Low-z, SDSS, SNLS and HST surveys.
This supernova Ia compilation uses The SALT 2 program to
transform light curves into distances using a modified version of
the Tripp formula \cite{tripp},
\beq
\mu = m_{B} - M + \alpha x_{1}-\beta c + \Delta_{M} + \Delta_{B},
\eeq
where $\mu$ is the distance modulus, $\Delta_{M}$ is a distance
correction based on the host-galaxy mass of the SNIa and $\Delta_{B}$
is the distance correction based on predicted bias from simulations.
Also, $\alpha$ is the coefficient of the relation between luminosity
and stretch; $\beta$ is the coefficient of the relation between
luminosity and color and $M$ is the absolute $B$-band magnitude
of the fiducial SNIa with $x_{1} = 0$ and $c = 0$. Also $c$ is the
color and $x_{1}$ is the light-curve shape parameter and $m_{B}$
is the log of the overall flux normalization. A covariance matrix
$\bf{C}$ is defined such that
\beq
\chi_{SNIa}^{2}
\, = \,
\Delta \vec{\mu}^{T}\cdot \mathbf{C}^{-1}\cdot \Delta \vec{\mu},
\eeq
where $\Delta \vec{\mu} = \vec{\mu}_{obs}- \vec{\mu}_{model}$
and $\vec{\mu}_{model}$ is a vector of distance modulus from a
given cosmological model and $\vec{\mu}_{obs}$ is a vector of
observational distance modulus. The $\vec{\mu}=\vec{m}-M$,
where $M$ is the absolute magnitude and $\vec{m}$ is the apparent
magnitude, which is is given by
\beq
\vec{m}_{model} \,=\, M+5Log_{10}\big(D_{L}\big)
\,+\, 5Log_{10} \Big(\frac{c/H_{0}}{1Mpc}\Big)
\,+\, 25 \,=\, \bar{M}+25+5Log \big(D_{L}\big),
\mbox{\quad}
\eeq
where $D_{L}=\frac{H_{0}}{c}d_{L}$ and
$\bar{M} = M+5Log(\frac{c/H_{0}}{1Mpc})$ is an nuisance parameter,
which depends on the Hubble constant $H_{0}$ and the absolute
magnitude $M$. To minimize with respect to the nuisance parameter
we follow a process similar to Refs.~\cite{conley,arjona}.
Therefore, the $\chi^{2}_{\bar{M} marg}$ is,
\beq
\chi^{2}_{\bar{M} marg}
\,=\,
\tilde{a}+ \log \Big( \frac{\tilde{e}}{2\pi}\Big)
- \frac{\tilde{b}^{2}}{\tilde{e}},
\eeq
where
\beq
\tilde{a}
= \Delta \vec{m}^{T} \cdot C^{-1} \cdot \Delta \vec{m},
\qquad
\tilde{b}
= \Delta \vec{m}^{T} \cdot C^{-1}\cdot \mathbb{I},
\qquad
\tilde{e} = \mathbb{I}^{T} \cdot C^{-1}\cdot \mathbb{I}.
\eeq
Here $\Delta \vec{m}= \vec{m}_{obs}-\vec{m}_{model}$
and $\mathbb{I}$ is the identity matrix.

\subsection{Power Spectrum}

The numerical analysis of the perturbations in our model can be
confronted with the power spectrum data of the BOSS-DR11 project
\cite{anderson}. For the comparison of these data with the theoretical
model described in the previous section, we use the chi-square
statistics,
\beq
\chi^{2}_{DR11} = \sum_{i=1}^{n=37}{\frac{\big[
P_{the}(z_{obs, i},k, \Omega^{0}_{dm},\Omega^{0}_{k},
b,\nu)-P_{obs,i} \big]^{2}}{\sigma^{2}_{obs,i}}},
\eeq
where $P$ denotes the power spectrum and we used the Planck
collaboration value $\Omega^{0}_{b}=0.049$
\cite{Planck2018Parameters}
and $H_{0}=70 \frac{km/s}{Mpc}$.
In our case, the theoretical power spectrum,
$(P_{the})$, results from the solution of the coupled system of
Eqs.~\eqref{38}, \eqref{39} and \eqref{41}-\eqref{43}. Namely, the
power spectrum of the normal (baryonic) matter at the current
redshift is determined as
\beq
P(k) = \big| \delta_{b}(k) \big|^2.
\eeq

Solving system of equations \eqref{38}, \eqref{39}, \eqref{41},
\eqref{42} and \eqref{43} requires specifying the initial conditions
for all the variables. For this end, we follow
Ref.~\cite{Fabris2012testing} and assume the transfer function
BBKS \cite{amvt, bardeen}, given by the expression
\beq
T(z) = \frac{\ln{(1+2.34 q)}}{2.34 q}
\left(1+3.89 q + 16.1 q^{2} + 5.64 q^{3} + 6.71 q^{4}\right)^{-1/4},
\eeq
where
\beq
q(k) = \frac{k}{h \Sigma \,Mpc^{-1}}
\eeq
and
\beq
\Sigma  \,= \, \Omega^{0}_{m} h \, \exp
\Big(-\Omega^{0}_{b}-\sqrt{2h}\frac{\Omega^{0}_{b}}{\Omega^{0}_{m}}\Big),
\eeq
where $\Omega^{0}_{m}=\Omega^{0}_{dm}+\Omega^{0}_{b}$
and $h=H_{0}/(100\,\,km{}\,s^{-1}\,Mpc^{-1})$.

\begin{table}\center
\begin{tabular}{|c|c|}
\hline
Parameters             & Best-fitting  \\ \hline \hline
$\Omega^{0}_{dm}$         &  $0.249 \pm 0.107$  \\
$\Omega^{0}_{k}$     &  $-0.05 \pm 0.120$     \\
$b$                    &  $0.000655 \pm 0.000300$    \\
$\nu$                   &  $0.000407 \pm 0.000070$    \\
\hline
\end{tabular}
\caption{\footnotesize Best-fitting parameters for $1\sigma$
confidence intervals and for the $\chi^{2}_{total}$ including
the SNIa and DR11 data sets.}\label{tab1}
\end{table}

The first results of the numerical analysis can be seen in Table 1.

\begin{figure}[t!]
\centering
\includegraphics[scale=0.6]{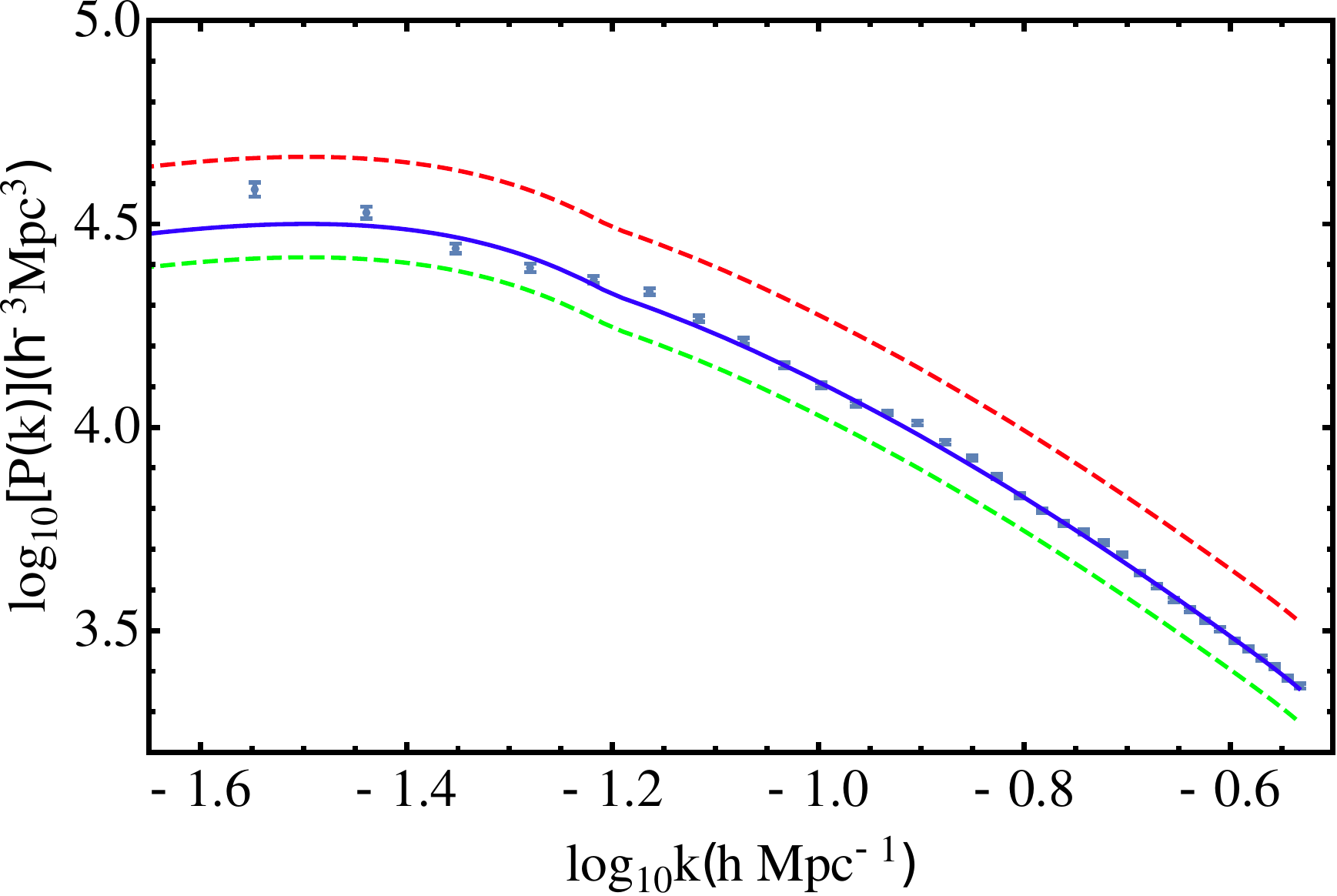}
\caption{\footnotesize Reconstruction of the power spectrum (PS)
of matter from the solution of the system of equations \eqref{38},
\eqref{39}, \eqref{41}, \eqref{42} and \eqref{43}. The blue line
is constructed with the best fitting from Table 1. The green and red dashed lines
correspond to the values of $b = 10^{-3}$ and $\nu = 10^{-3}$, respectively, letting fixed all the other values. It is evident that the matter
PS is quite sensible to $\nu$ and $b$ values. On the other hand,
no essential changes under variation of $\Omega^{0}_{k}$ were
observed.}
\label{fig2}
\end{figure}

The plot for the power spectrum which results from the approach
explained above, is shown in Figure \ref{fig2}. The blue line is
calculated using the best fitting given by the Table \ref{tab1}. The
other lines keep all the parameters fixed while varying the parameter
of the running $\nu$ (red line) or the warmness $b$ (green line).

\begin{figure}
 	\centering
	\includegraphics[scale=0.440]{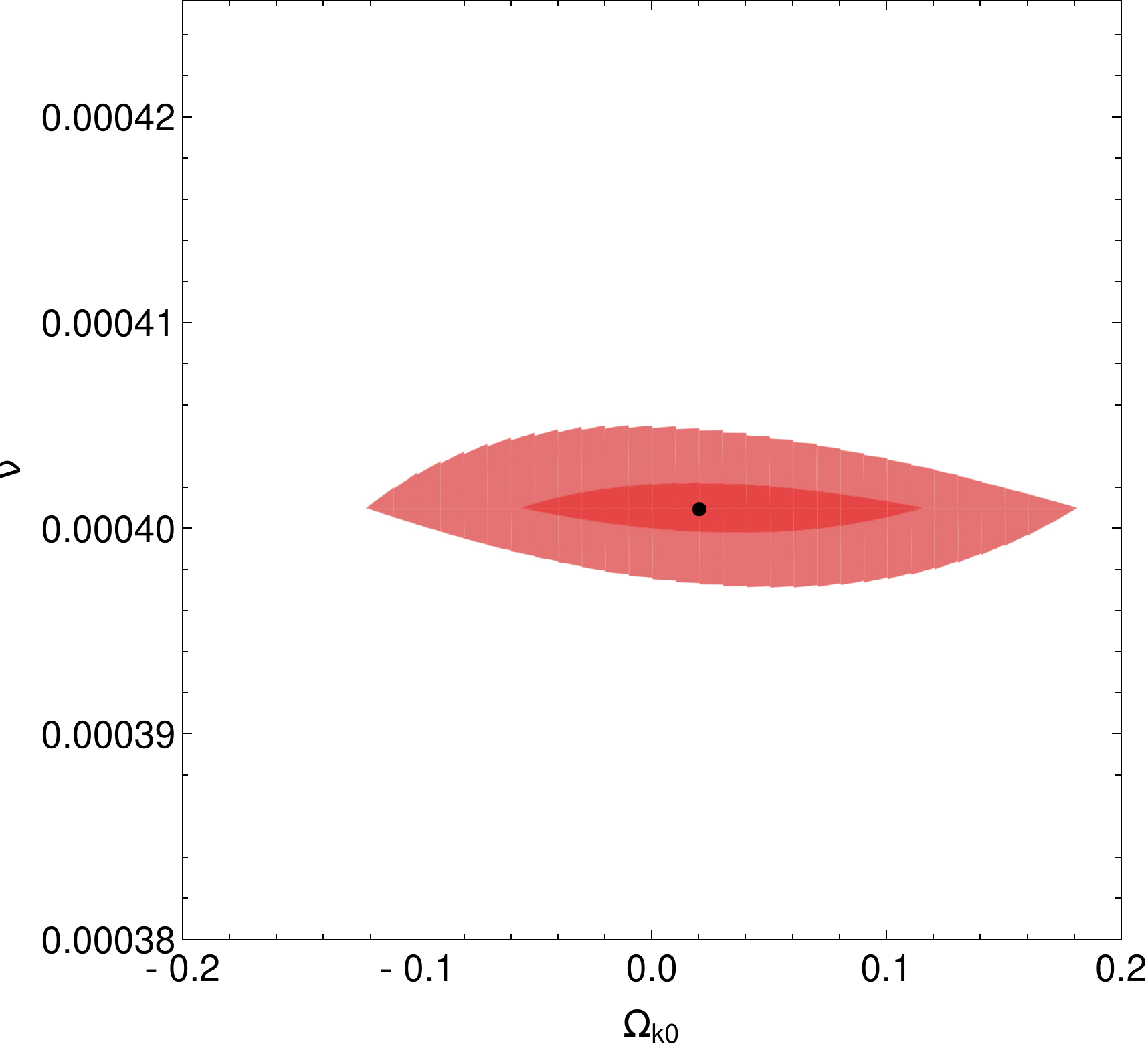}
	\includegraphics[scale=0.440]{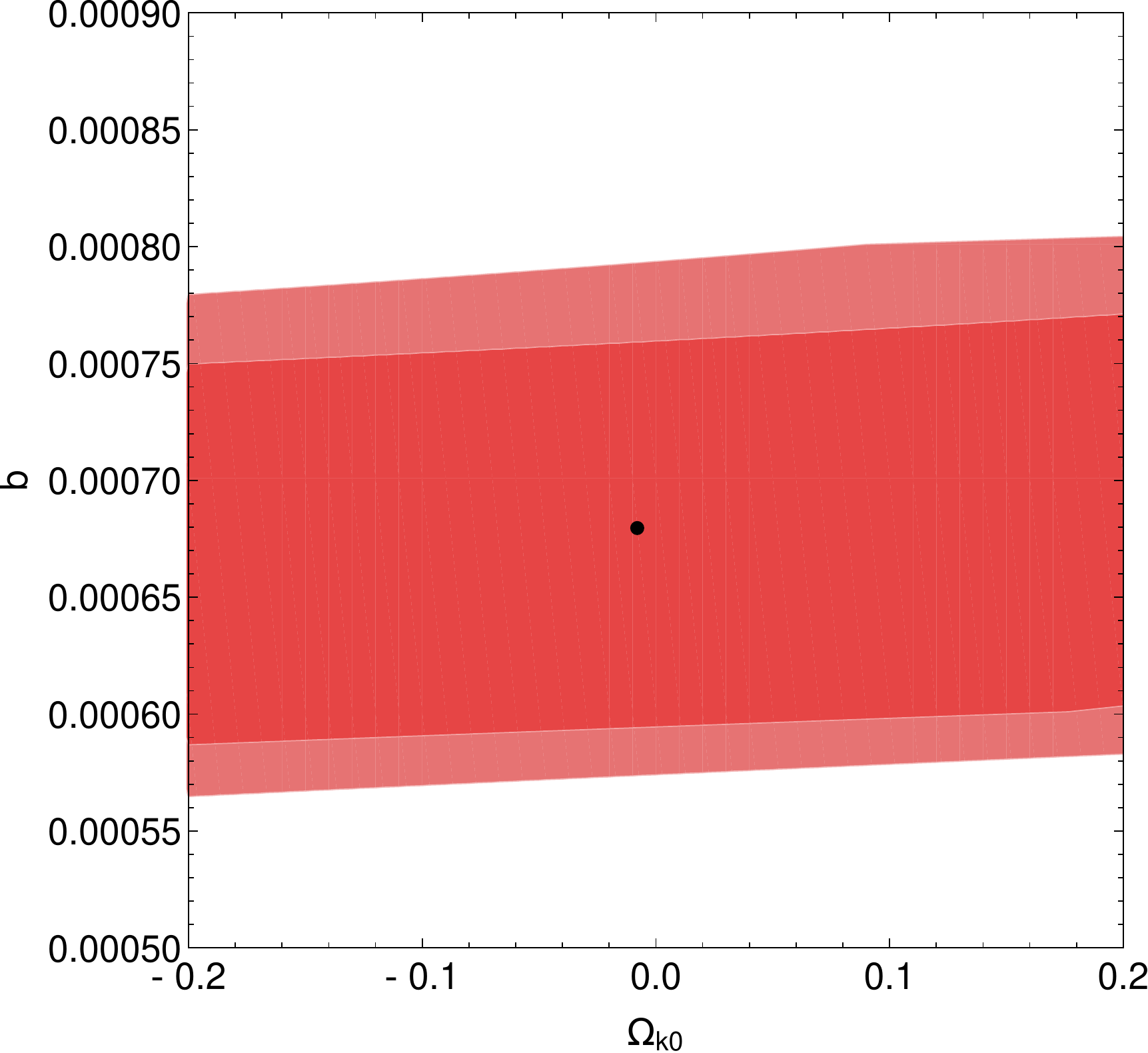}
	\includegraphics[scale=0.440]{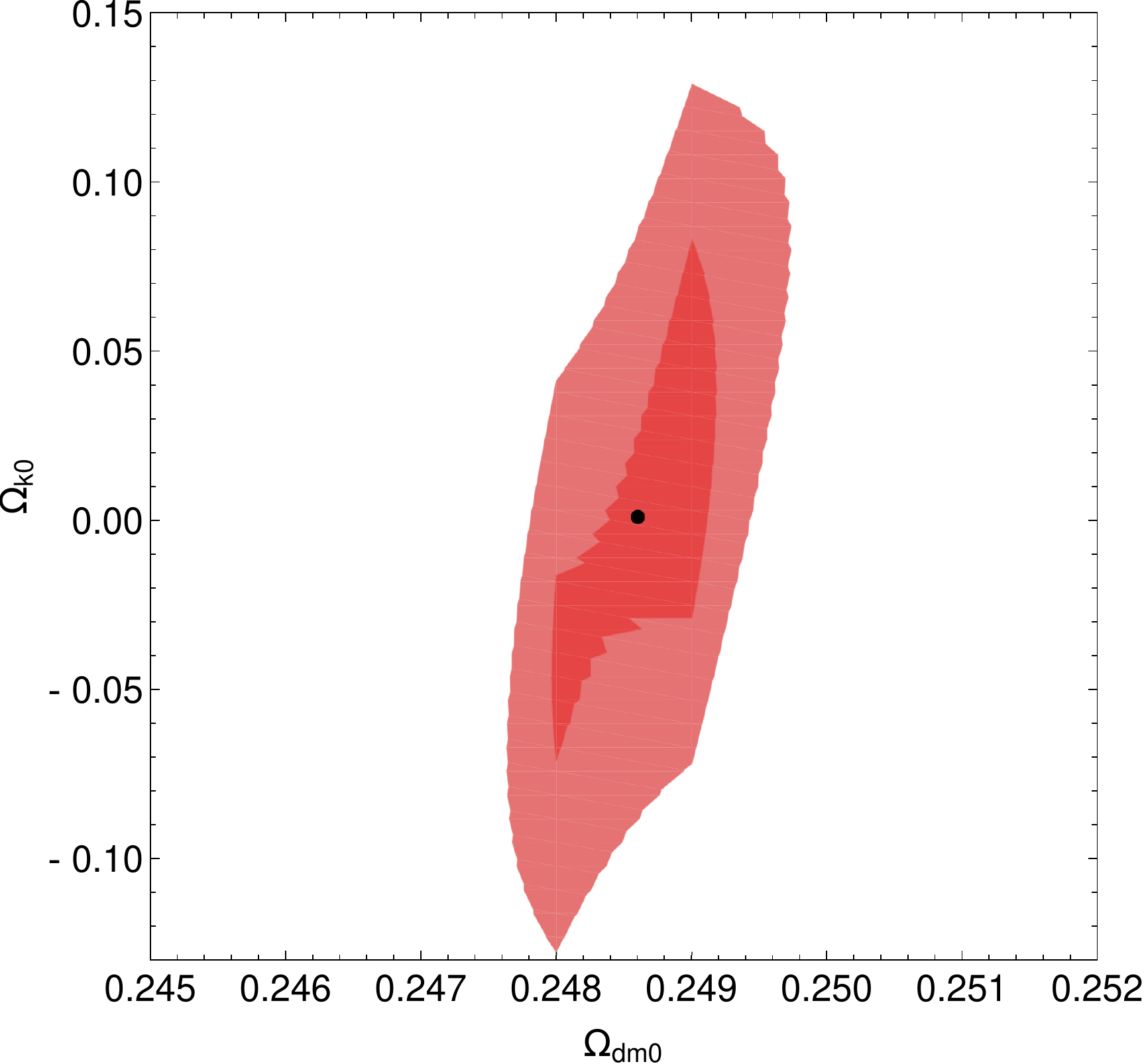}
 	\caption{\footnotesize
 Observational constraints of the curvature parameter $\Om^0_k$
 versus the free parameters of the model
($\Om^0_{dm}$, $b$, $\nu$), based on $\chi^{2}_{total}$. In the
top figure on the right we show the degeneracy on $\Om^0_k$
versus $b$. In the top figure on the left side there are constraints
with respect to parameter $\nu$ and at the bottom we show constraints
with respect to parameter $\Omega^{0}_{dm}$. In all cases we assume
$H_0=70 \frac{km/s}{Mpc}$
and $ \Omega^{0}_{b} = 0.049$.}
\label{fig3}
\end{figure}

To determine the observational constraints using the two data sets
described above, namely SNIa and matter power spectrum DR11, we
define the total $\chi^2$ as the sum of the individual contributions,
in the form
\beq
\chi^{2}_{total} = \chi^{2}_{SNIa}+\chi^{2}_{DR11}
\eeq
and elaborate this value using the data presented in this section.
The results of this treatment are illustrated in Fig.~\ref{fig3},
where we  show how the parameters that characterize our model vary
with respect to the curvature $\Omega^0_k$. One can observe that
the two data sets are complementary, except the case of the plane
$(\Om^0_k, b)$.

\section{Conclusions}
\label{s5}

We have developed a cosmological model with a non-zero spatial
curvature, running cosmological constant and warm Dark Matter
(WDM). The WDM is described in terms of the RRG model and
the form of the hypothetical running of $\rho_\La$ is fixed by the
arguments of covariance of the effective action of vacuum.
The ideal fluid describing normal matter assume a general equation of
state with a constant $\om$. At the background level, we have found
the analytical and general expressions for the corresponding relative
energy densities, as well as the expansion rate given by the Hubble
parameter.

The analysis of density perturbations for all involved fluids leads
to the system of equations for the density contrasts. The new element
compared to the previous works
\cite{CCwave,Fabris2012testing,RGGRcosm} is that this time we
took into account modifications caused by the spatial curvature,
 including on the expansion rate. This system of equations has been
 solved numerically to reconstruct the corresponding matter power
 spectrum and to impose the restrictions on the free parameters of
 our model, such as the running parameter $\nu$ and the warmness
 $b$, taking into account the effect of the spatial curvature
 $\Om^0_k$.

Once we include spatial curvature, the observational constraints
restrict the magnitudes of both parameters $\nu$ and $b$ at the
order of $10^{-4}$. The combined data prefers a relatively low
warm dark matter component of the order of $ \Om^0_{dm} = 0.25$,
as a component of the total matter balance today,
$\Om^0_m = \Om^0_{dm} + \Om^0_b = 0.299$. However, there is a
degeneracy in the observational constraints
on the parameter $\Om^0_ {k}$, which can be clearly observed
in the diagram showing the plane $(\Omega^{0}_{k},b)$ in
Fig.~\ref{fig3}. Even though, a slight preference for a closed
universe can be identified.

In conclusions, adding the curvature parameter to our model of
running cosmological constant with WDM, increases the number
of dimensions of the parameter space, regardless the effect of
space curvature is phenomenologically not very strong.
Additional tests may be useful to obtain more robust constraints
on the parameters  $\nu$ and $b$. In particular, we expact that
the observational constraints coming from CMB and BAO may
give a strong enforcements of our results. The theoretical basis
of these tests would be a natural continuation of the present work.

On the other hand, considering the generality of the model and
its ability to describe different phases of the Universe, another
possible perspective for further investigations work could be to
study the Hubble tension by including the parameter $w$, defined
in Eq.~(\ref{6}), as a new free parameter and estimating the
Hubble parameter today. We consider this possibility for a
possible future works.

\section*{Acknowledgments}

\noindent
J. A. Agudelo Ruiz thanks CAPES for supporting his PhD project.
J. C. Fabris thanks  Fundação de Amparo à Pesquisa e Inovação do
Espírito Santo (FAPES, project number 80598935/17) and Conselho
Nacional de Desenvolvimento Cient\'{i}fico e Tecnol\'{o}gico (CNPq,
grant number 304521/2015-9) for partial support. This work of I.Sh.
was partially supported by CNPq under the grant 303635/2018-5.





\end{document}